\documentclass[11pt,twoside]{article}

\usepackage{asp2014}

\aspSuppressVolSlug
\resetcounters

\begin{document}

\title{GDL 1.1, a smart and green language}

\author{A.~Coulais,$^{1,2}$ and G.~Duvert$^3$
  \affil{$^1$LERMA, Observatoire de Paris, Universit\'e PSL, Sorbonne Universit\'e,
    CNRS, Paris, France \email{Alain.Coulais@obspm.fr}\\
    $^2$Universit\'e Paris-Saclay, Universit\'e Paris Cit\'e, CEA, CNRS, AIM, Gif-sur-Yvette, France\\
    $^3$IPAG, CNRS, Universit\'e de Grenoble Alpes, Grenoble, France}
  }


\paperauthor{A. Coulais}{Alain.Coulais@obspm.fr}{0000-0001-6492-7719}
            {Observatoire de Paris}{LERMA}{Paris}{}{75014}{France}

\paperauthor{G. Duvert}{gilles.duvert@univ-grenoble-alpes.fr}{0000-0001-8769-3660}
            {OSUG}{}{Saint-Martin-d'Heres}{Rh\^one-Alpes}{38400}{France}



\begin{abstract}

GDL, a free interpreter for the IDL language, continues to develop
smoothly, driven by feedback and requests from an increasingly active
and growing user base, especially since GDL was made available on
GitHub. Among the most notable features introduced in recent years are
stable Widgets; extensive testing on M1, M2, and M3 processors; excellent
computational performance (including OpenMP support) demonstrated
across a comprehensive benchmark; simplified compilation and
installation processes; and the availability of SHMMAP and Bridge
functions, which enable concurrent GDL runs on shared RAM in HPC
environments.

As developers of GDL, we believe this language holds a valuable place
in today's world, where efficiency and low-power computing are
essential. GDL (not to mention IDL), written in C/C++, demonstrates
exceptional efficiency in "real-world" benchmarks, making it one of
the few interpreted languages that can truly be considered "green."
Moreover, it is likely the only interpreter accompanied by a vast
collection of free, well-tested, and proven astronomical procedures
developed by colleagues over the years. GDL also stands out for its
suitability for long-term projects, thanks to its stable and reliable
syntax.

\end{abstract}

\section{What is GDL ? A free clone of IDL !}

Despite being created in the mid-1970s, IDL remains a modern
programming language with several powerful features. It boasts a
smart, concise, and efficient syntax and comes with a extensive set of
fast internal procedures. These resources provide direct access to a
wide range of functionalities, including advanced plotting
capabilities, support for large data input/output, and native
multi-core computations.

Over decades, a vast collection of scientific libraries and pipelines
has been written in IDL, highlighting its enduring relevance in the
scientific community. As shown in Fig. 1 in \citet{Astropy2022}, IDL
is still the second most widely used programming language in astronomy
today.

GDL \citep{Park2022} is a free and open-source interpreter of the IDL
language. It can interpret IDL syntax in version 8, and comes with a
very large set of internal functionnalities, including all the core
functions of IDL.  GDL is continuing smoothly its development, driven
by feedback's and requests from end-users who are more numerous and
active now that GDL is hosted on GitHub.
(\url{https://github.com/gnudatalanguage/gdl})
GDL run on Linux, *BSD,
MS-win and OSX, on processors x86\_64, ppc64le, s390x and aarch64
(ARM, M1/2/3 included).

\section{Why GDL in 2024 ?}

The GDL project was driven by the need to save all the knowledge
amassed in the last 40 years in the million of lines of IDL code used
in so many parts of astronomy and space sciences up to now
--noticeably a lot of data reduction pipelines-- should its costly
closed-source counterpart IDL disappear. It was extensively discuted
in \citet{CoulaisADASS2012}.
But GDL has proven since to be useful for day-to-day research, is
known for its excellent publication-quality graphics, and is as
efficient and ``green'' than an interpreted language can be.
Furthermore, due to its opensourceness, GDL fulfills the modern
requirements for freely reproducible research (FAIR principles),
something IDL cannot claim.

\section{What is new recently ?}

GDL 1.1 was delivered in November 2024.
Among the most interesting features introduced during the last year
are the large test on M1/M2/M3 processors (running OSX or Linux), the
very good performance for computation (OpenMP included) demonstrated
on a wide benchmark (for ARM \& x86\_64), a simplification of
compilation and installation, the availability of SHMMAP and Bridge
functions that are great for deploying concurrent GDL runs on the same
RAM on HPC.

\section{Low carbon language !}

As developers of GDL we believe that this language has a useful place
in the modern world where efficiency and low power computing is a
must. Indeed, GDL (not to mention IDL), written in C/C++ shows
excellent efficiency in "real world" benchmarks, making it one of a
few interpreted language that are really "green" \citep{Energy2017}. And
probably the only one that comes with such a huge collection of free,
multi-tested and proven astronomical procedures written by colleagues
in the past.

One reason why GDL can be faster than IDL in several places is that it
uses modern optimized routines, published under Open Source licenses,
that are forbidden to IDL. For example: Eigen::, FFTw3, dSFMT for
random number generation, delaunator for triangulation, two ultra fast
median filtering codes\ldots

GDL is also interesting for long life code since it is based on a very
stable syntax \citep{CoulaisADASS2012} now available in ANTLR4.

Reducing code size and avoiding recode is also {\em green}. IDL syntax
is particularly concise. Written in C \& C++ the core code of GDL is
{\em only} 225 k-lines. This because GDL uses a large set of
open-source third party libraries, mostly under GNU GPL v2/3 or BSD
licenses. The use of external libraries reduce the code we have to
maintain and we benefit from progress made by experts in others
fields (algorithms, compilers, internal tricks in processors ...)

And finally, for large computations, GDL is not hindered by the
``genial'' Global Interpretor Lock (GIL) present in Python.

\section{GDL for end users}

\subsection{GDL community}

We do have a thread for discussions in the GitHub interface.  We do have
also a diffusion list reporting new releases and few critical bugs.
For advanced users, we would recommend to register to receive every new
issue submitted in GitHub. Bugs reports (issues) are always scrutinized.

\subsection{Scientific Formats and Save files}

As a free and open-source substitute for IDL and PV\_WAVE, GDL is
able to read and write major scientific formats like FITS, HDF and
HDF5, NetCDF... but also PDS (Planetary Data System), GeoTIFF, GRIB,
DICOM formats. Plus all common image formats (GIF, JPEG, TIFF..).
Data in basic text or CVS files can be read in one-line command.
GDL is fully compatible with IDL ``save'' (XDR) files both in reading and
writing, with very good performances (even for very large files). In
this release it also allows to save and restore (GDL only) compiled
procedures and functions.

\subsection{Binding with C/C++ codes}

If you would like to incorporate in GDL a program you wrote in C/C++,
it is super easy. Several solutions exists. The classical one
is to use {\tt call\_external}, which exist since the beginning of GDL and
was widely used to process HFI Planck raw data.  Another one is to
declare in few lines an interface to C/C++ codes.  This is the way we
use to call GSL, Eigen::, FFTw codes ...

\subsection{Extensions to the CLI}

IDL CLI (command line interface) uses a very limited sub-set of shell
short-cuts (ctrl-a, ctrl-e ...). Thanks to the use of Readline we
extended it in GDL to most of basic shortcuts existing in
Bash/Emacs. A list is available by typing {\tt GDL>
  help,/all\_keys}. Auto-completion of input files names is also
available using the Tab.

We also add the {\tt\#} operation, which is a shortcut to see quickly at the
CLI level various information about a function (resp. procedure) : if the
function is internal, if not if the function is already compiled, and
the way to call it. E.g.:\\
{\footnotesize {\tt GDL> \#beselj \\
Internal FUNCTION : res=BESELJ([2 Args],DOUBLE,ITER,HELP)\\
GDL> \#dist\\
No Procedure/Function, internal or compiled, with name : DIST\\
GDL> .r dist\\
GDL> \#dist\\
Compiled FUNCTION : res=DIST([2 Args])}}

\subsection{Smart language ? Fast writing of proof of concept !}

As shown in \citet{GastaudADASS2024} it was very easy to quickly
develop in GDL a very efficient pipeline as an alternative to the
official one to process MIRI data observed by JWST.  Reading FITS
files was done using the routines available in IDL Astro lib
\url{https://github.com/wlandsman/IDLAstro}.
Revisited ramp processing
was done by a novel algorithm using SHIFT and MEDIAN operations, and
statistical was checked using Allan variance.  Spatial and temporal
deglitching was done with the very efficient MEDIAN tool we
have. Odd/Even flat field correction was also done with basic
iterative mixing of primitives functions available in GDL.
Re-projection of the mosaics was done with basic SHIFT, REBIN. No new
code developed, just usage of a series of basic bricks containing high
levels procedures. No need to {\em enter} into very complicated
infrastructure as is the official JWST pipeline.

This very simple and elegant usage of GDL is a clear illustration of
the quality of IDL concepts. Just tens of lines of codes to write an
alternative pipeline which runs hundred of time faster than the python
one, and was enough to got better results, already published \citep{Bouchet2024}.

\section{Future of GDL}

Since about the Planck ``era'' \citep{CoulaisADASS2014} when GDL was
widely used to process Planck HFI data, the core code of GDL is very
stable and widely tested. We regularly update the code inside to stick
with progress of the multi-cores CPU (OpenMP; Eigen:: was a good
choice !) and as said above we use all modern optimized algorithms we can
find in the open source community.

With a growing and more active community of users gathered on GitHub,
GDL is now routinely facilitating the transition from IDL to GDL for
numerous existing scientific codes, including the SolarSoftWare (SSW)
framework. We hope to encourage more people to contribute, even in
small ways, to the ongoing improvement of GDL, in particular in
enhancing the coverage of the core code with unit tests.

In GDL, you don't need to optimize your code, it is already done,
since the internal code is already as fast as possible and is not
stuck by the GIL~!  Just concentrate on the ideas and the concepts.
Extra bonus : GDL is FAIR and green.

\acknowledgements

AC sincerely thanks the packagers and also users who report bugs via the
GDL GitHub issues page. Despite extensive unit tests, regressions may
still occur. AC also appreciates the Scientific Council at the Paris
Observatory for supporting the acquisition of a shared M2 computer
running OSX, POL (and CEA) for travel funding, and RG for their
long-standing use of GDL.
 
\bibliography{ADASS2024_coulais}

\begin{thebibliography}{}
\expandafter\ifx\csname natexlab\endcsname\relax\def\natexlab#1{#1}\fi
\expandafter\ifx\csname url\endcsname\relax
  \def\url#1{\texttt{#1}}\fi
\expandafter\ifx\csname urlprefix\endcsname\relax\def\urlprefix{URL }\fi
\providecommand{\eprint}[2][]{\url{#2}}

\bibitem[{Bouchet \& {al.}(2024)}]{Bouchet2024}
Bouchet, P., \& {al.} 2024, The Astrophysical Journal, 965, 51.
  \urlprefix\url{https://dx.doi.org/10.3847/1538-4357/ad2770}

\bibitem[{{Coulais} \& {al.}(2012)}]{CoulaisADASS2012}
{Coulais}, A., \& {al.} 2012, in Astronomical Data Analysis Software and
  Systems XXI, edited by P.~{Ballester}, D.~{Egret}, \& N.~P.~F. {Lorente},
  vol. 461 of Astronomical Society of the Pacific Conference Series, 615

\bibitem[{{Coulais} \& {al.}(2014)}]{CoulaisADASS2014}
--- 2014, in Astronomical Data Analysis Software and Systems XXIII, edited by
  N.~{Manset}, \& P.~{Forshay}, vol. 485 of Astronomical Society of the Pacific
  Conference Series, 331

\bibitem[{Gastaud \& Coulais(2024)}]{GastaudADASS2024}
Gastaud, R., \& Coulais, A. 2024, in Astronomical Data Analysis Software and
  Systems XXXIV, Astronomical Society of the Pacific Conference Series

\bibitem[{Park \& {et al}(2022)}]{Park2022}
Park, J., \& {et al} 2022, Journal of Open Source Software, 7, 4633.
  \urlprefix\url{https://doi.org/10.21105/joss.04633}

\bibitem[{Pereira \& {al.}(2017)}]{Energy2017}
Pereira, R., \& {al.} 2017, in Proceedings of the 10th ACM SIGPLAN
  International Conference on Software Language Engineering (New York, NY, USA:
  Association for Computing Machinery), SLE 2017, 256–267.
  \urlprefix\url{https://doi.org/10.1145/3136014.3136031}

\bibitem[{{The~Astropy~Collaboration}(2022)}]{Astropy2022}
{The~Astropy~Collaboration} 2022, The Astrophysical Journal, 935, 167.
  \urlprefix\url{http://dx.doi.org/10.3847/1538-4357/ac7c74}

\end{thebibliography}

\end{document}